\documentclass[reprint, amsmath, amssymb, aps, showpacs, prl]{revtex4-2}

\usepackage{graphicx}
\usepackage{dcolumn}
\usepackage{bm}
\usepackage{hyperref}
\usepackage{physics}

\begin{document}
\title{Persistent currents in rings of ultracold fermionic atoms}

\author{Yanping Cai}
\author{Daniel G. Allman}
\author{Parth Sabharwal}
\author{Kevin C. Wright}

\email{Kevin.Wright@Dartmouth.edu}

\affiliation{Department of Physics and Astronomy, Dartmouth College, 6127 Wilder Laboratory, Hanover NH 03755, USA}

\begin{abstract} 

We have produced persistent currents of ultracold fermionic atoms trapped in a ring, with lifetimes greater than 10 seconds in the strongly-interacting regime. These currents remain stable well into the BCS regime at sufficiently low temperature. We drive a circulating BCS superfluid into the normal phase and back by changing the interaction strength and find that the probability for quantized superflow to reappear is remarkably insensitive to the time spent in the normal phase and the minimum interaction strength. After ruling out spontaneous current formation for our experimental conditions, we argue that the reappearance of superflow is due to weak damping of normal currents in this limit. These results establish that ultracold fermionic atoms with tunable interactions can be used to create matter-wave circuits similar to those previously created with weakly-interacting bosonic atoms.
\end{abstract}

\maketitle

Progress in understanding quantum fluids has often been made by considering spherical, cylindrical, toroidal, or more exotic geometries~\cite{Fradkin2013}, and circuits built from quantum materials have many important applications including quantum computing. Quantum gases with periodic boundary conditions provide unique opportunities for exploring quantum many-body physics, especially where it is possible to bias a circuit with an external flux~\cite{Amico2005, Pecci2021}. One crucial characteristic of such circuits is that they can support quantized currents that flow without being driven by an external power source. Persistent non-equilibrium currents are commonly understood to occur in superconducting~\cite{Deaver1961} and superfluid~\cite{Bendt1962} phases, but equilibrium persistent currents can also appear in normal conducting phases around closed paths shorter than the coherence length~\cite{ButtikerJosephsonPLA1983, Bluhm2009, Bleszynski-Jayich2009}. The current response of such circuits to external flux often conveys important information about the state of the system~\cite{Little-ParksPRL1962}.
 
Previous experiments with multiply-connected ultracold gases have utilized weakly-interacting atomic Bose-Einstein condensates (BEC) in magnetic and optical traps~\cite{GuptaBose-EinsteinPRL05, RyuObservationPRL07, Bruce2011, BeattiePersistentPRL13, Neely2013, SherlockTimePRA11, Navez2016, DeGoerDeHerve2021}. Experiments on ring BECs have demonstrated the existence of metastable currents~\cite{RyuObservationPRL07, BeattiePersistentPRL13} and quantized phase slips~\cite{WrightDrivingPRL13, RamanathanSuperflowPRL11}. Bosonic superfluid circuits have been constructed by incorporating Josephson junctions~\cite{RyuExperimentalPRL13, EckelHysteresisN14, EckelInterferometricPRX14}, and the experimental usefulness of multiply-connected quantum gases has been demonstrated by studies of collective-mode precession~\cite{Marti2015}, spontaneous currents~\cite{WeilerSpontaneousN08, Corman2014, Aidelsburger2017}, quantum turbulence~\cite{Neely2013},  propagation of shock waves~\cite{Wang2015}, the stability of supersonic superfluid flows~\cite{Pandey2019, Guo2020}, and more~\cite{Amico2021}.

Fermionic quantum gases provide access to a rich variety of physics distinctly different from that of purely bosonic systems. Furthermore, magnetic fields can often be used to continuously tune interactions between fermionic atoms from the weakly attractive limit where BCS pairing can occur to the weakly repulsive limit where the atoms can form a BEC of weakly-bound molecules. Fermionic superfluidity has been extensively studied throughout this BEC-BCS crossover, including experimental observations of an interaction-dependent critical velocity in 3D~\cite{Weimer2015}, and very recently in 2D~\cite{Sobirey2021}. Josephson junctions and quantum point contacts have also been realized in singly-connected fermionic quantum gases~\cite{Valtolina2015, Husmann2015}. 

In this work we report the first creation of a multiply-connected superfluid ``circuit'' in an ultracold Fermi gas, and show that it is possible to reliably create and detect quantized currents in this system. We demonstrate that currents can survive well into the fragile BCS regime and examine the decay and revival of currents after quenches to the normal phase in this limit, establishing a foundation for other proposed experiments involving quantum gases in rings and ring lattices~\cite{Amico2005, Girardeau2008, Yanase2009, Roncaglia2011, Edmonds2013, Aghamalyan2015, Ragole2016, Metcalf2017,  Pecci2021}.

In these experiments, we use a quantum degenerate gas of $^6$Li atoms in an equal mixture of the lowest-energy spin states ($\ket{m_J = -1/2,  m_I = 1}$ and $\ket{m_J = -1/2,  m_I = 0}$). Interactions between atoms in these two states are attractive (repulsive) at magnetic fields above (below) a broad ($>10$ mT) Feshbach resonance at 83.2 mT. To create a persistent current, we must have a continuous (pair) superfluid around a closed path, which requires a trap with a smooth ring-shaped potential minimum and cooling the system below a critical temperature which depends on the interaction strength and density. We created an optical ring trap with two red-detuned laser beams, a horizontal ``sheet'' beam ($\lambda=1068$ nm,  horizontal waist 290 $\mu$m, vertical waist 7 $\mu$m) and a vertical ring-pattern beam ($\lambda$ = 780 nm, average radius 12.0(1) $\mu$m, radial $1/e^2$ half-width 2.2(1) $\mu$m (see Supplemental Materials). 

\begin{figure}[t!]
	\centering
	\includegraphics[width=\columnwidth]{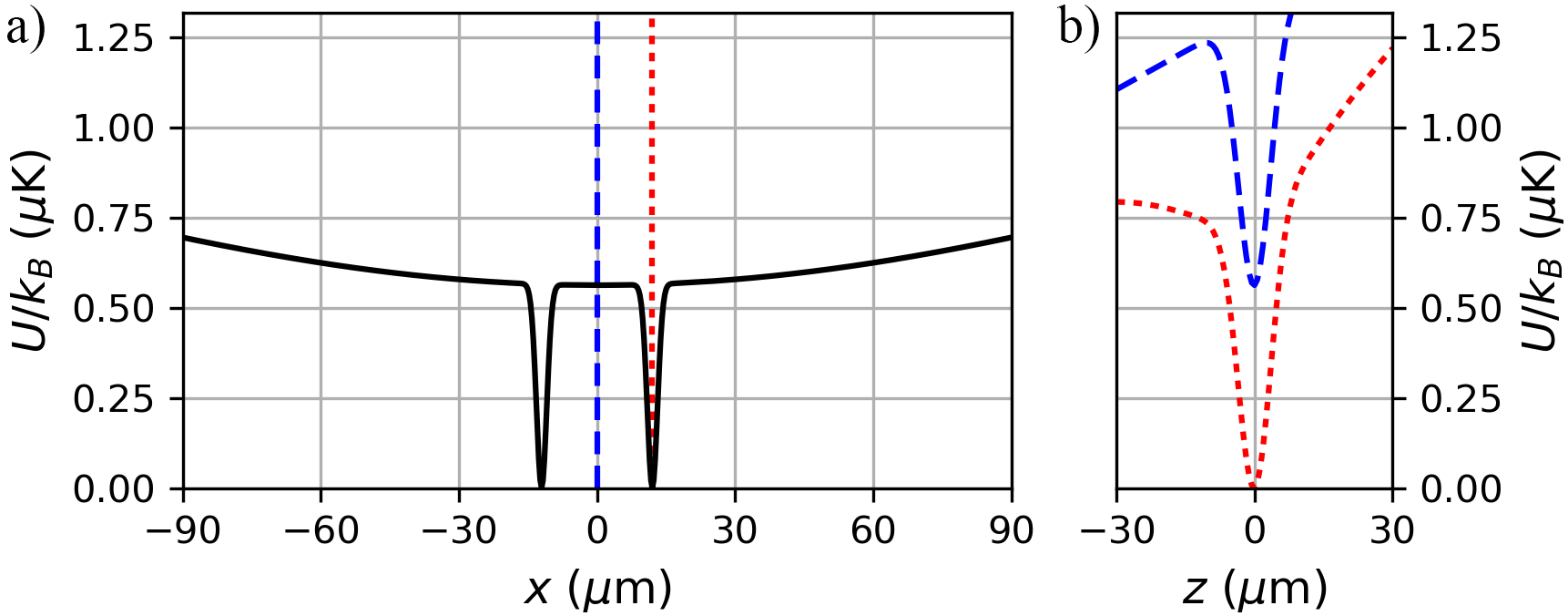}
	\caption{ Cross sections of an idealized model of the potential experienced by a $^6$Li atom in our ``ring-dimple'' optical trap in its final configuration (the potential is twice as deep for molecules). (a) Black line is the potential along a horizontal line through the center of symmetry, transverse to the direction of propagation of the ``sheet'' beam. The radial trap frequency for atoms away from the ring minimum is 37(5) s$^{-1}$.  b) Vertical cross section of the trap potential at $r=0$ (dashed line, blue online) and $r=12.0$ $\mu$m (dotted line, red online), as indicated by corresponding vertical lines in (a). The vertical trap frequency is $1.5(1)\times10^3$ s$^{-1}$ for atoms near the ring potential minimum, and $1.4(1)\times10^3$ s$^{-1}$ away from the ring beam.The plot vertical range is from $U_\text{trap}=0$ at the ring minimum to the ``trap-off'' potential in the midplane of the ring (1.32 $\mu$K).}
	\label{fig:trap_potential}
\end{figure}

The next critical requirement is to achieve and maintain low enough temperatures to study supercurrents over a wide range of interaction strengths. We loaded the atoms into the ring trap with the sheet beam initially at high power (4 W), and performed final evaporative cooling with the magnetic field near resonance (82.0 mT) by decreasing the sheet beam power to 40 mW while holding the ring beam power at 0.85 mW. Evaporation occurred as molecules fell out of the bottom of the ring region where the potential barrier was lowest; the gravitational gradient reduced the final evaporation depth to an estimated $k_B\cdot0.80(5)$ $\mu$K as shown in Fig.~\ref{fig:trap_potential}(b).

After evaporation there were $1.0(1)\times10^4$ atoms in each spin state, paired into weakly-bound molecules with strong repulsive interactions. The chemical potential ($\mu$) was high enough that most of the molecules were not localized to the ring and formed a wide, thin disk in the radially weak, vertically strong harmonic potential of the sheet beam ($\nu_r=37(5)$ s$^{-1}$, $\nu_z=1.4(1)\times10^3$ s$^{-1}$). The fraction of the population in this ``halo'' increased if we subsequently tuned interactions to the weakly attractive (BCS) limit where $\mu\approx E_F$ (Fermi energy). From a model of our trap we calculated that in this limit $E_F=h\cdot16(1)\times10^3$ s$^{-1}=k_B\cdot0.77(6)$ $\mu$K (see Supplemental Materials). The radial trap frequency for atoms (and molecules) near the ring potential minimum was $\nu_r=4.0(2)\times10^3$ s$^{-1}$.

\begin{figure}[t!]
	\centering
	\includegraphics[width=\columnwidth]{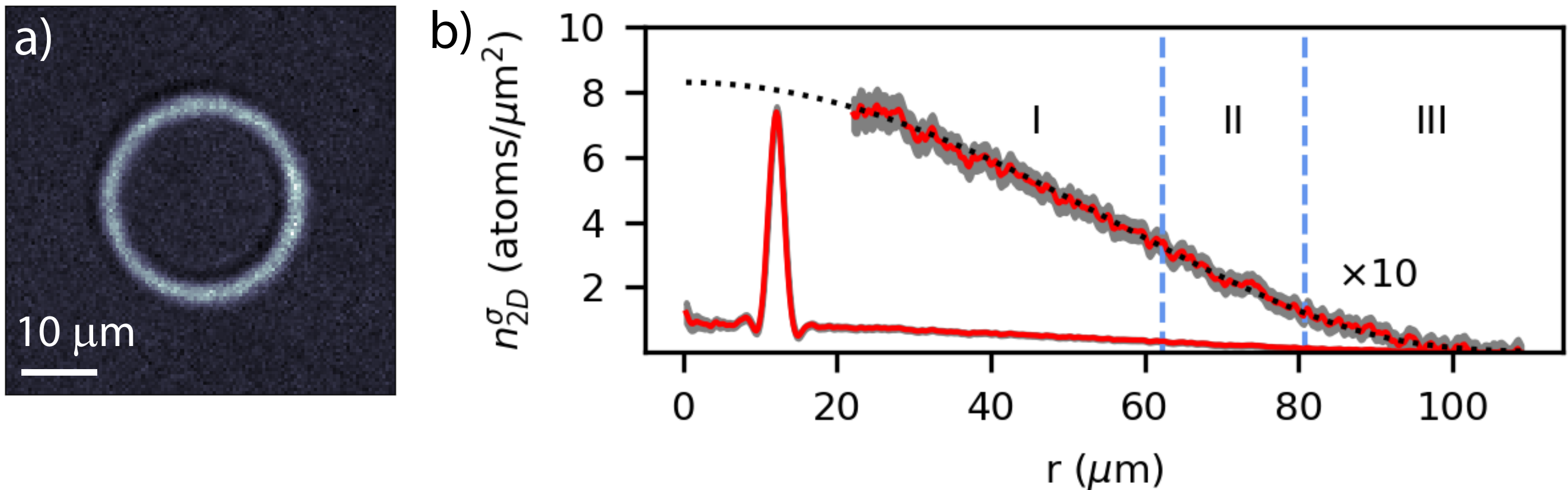}
	\caption{ a) Absorption image (10 averaged) of an equal spin mixture of $^6$Li atoms in the trap potential of Fig.~\ref{fig:trap_potential}, with $1\times10^4$ atoms in each spin state. A magnetic field of 107.4 mT has been used to tune the scattering length to $-179$ nm. The density peak is at $r=12.0(2)$ $\mu$m. This is a 50 $\times$ 50 $\mu$m region cropped from a 215 $\times$ 215 $\mu$m image. (b) Radial column density obtained by azimuthal averaging over the full field of view. The plot is shown vertically rescaled $\times10$ for $r>20$ $\mu$m to emphasize the broad halo extending to $r=100$ $\mu$m. The black dotted line is the expected density profile ($\times$10) for an ideal Fermi gas in our trap at $T= 25$ nK. In region I the system is 3D degenerate, II is quasi-2D degenerate, and III is quasi-2D thermal.  Grey band: $2 \sigma$ variation of $n_{\text{2D}}(r)$ when calculated separately for each image in the set.}
	\label{fig:ring_insitu}
\end{figure}

The low-density halo was hardly visible in absorption images, but easily observed in radial plots of the column density after azimuthal averaging. Fig.~\ref{fig:ring_insitu} shows averaged results from 10 runs where the field was ramped from 82.0 mT to 107.4 mT before imaging the atoms in the ring. Fig.~\ref{fig:ring_insitu}(a) has been cropped to show ring density variations (10\% peak-to-peak) in more detail. Fig.~\ref{fig:ring_insitu}(b) shows the column density,  $n_{\text{2D}}(r)$, obtained from the full-frame image by averaging data in radial bins 1 $\mu$m wide. Fitting the density profile for $r>20$ $\mu$m with a model of an ideal Fermi gas in our trap potential indicated that $T=25(5)$ nK. Because $T< h \nu_z/k_B$ = 72 nK, we accounted for the crossover from 3D to quasi-2D in the outer regions of the halo (see Supplemental Materials). The best fit of this model to the data for $r>20$ $\mu$m is shown in Fig.~\ref{fig:ring_insitu}(b) as a dotted black line. 

When we ramped from 82.0 mT  to 68.0 mT (BEC regime) we found that $n_{\text{2D}}$ for $r>20$ $\mu m$ had the Gaussian profile expected for a thermal gas of molecules at 90(3) nK. Temperature changes are expected for isentropic interaction ramps because the temperature dependence of the entropy is different in the BEC and BCS limits~\cite{Carr2004}. The minimum temperature we observed in the BCS regime was likely limited by heating due to hole creation by collisions with background gas molecules~\cite{TimmermansPRL01}. Fermionic systems are especially sensitive to heating at low temperatures. Retaining a large part of the population in a low-density halo increases the average heat capacity per particle and reduces the heating rate, which was important for the experiments described below.

Our general procedure for creating persistent currents and studying their stability in the BCS regime was the following: We prepared a strongly-interacting molecular BEC at $B=82.0$ mT as described above, then initialized the current state by stirring with a blue-detuned laser beam that created a localized repulsive potential. We then changed the interaction strength adiabatically by ramping the magnetic field up to the BCS regime, then ramped back to 82.0 mT. Finally, we ramped to the BEC regime and used a self-interference technique to determine the final current state of the ring. 

\begin{figure}[t!]
	\centering
	\includegraphics[width=\columnwidth]{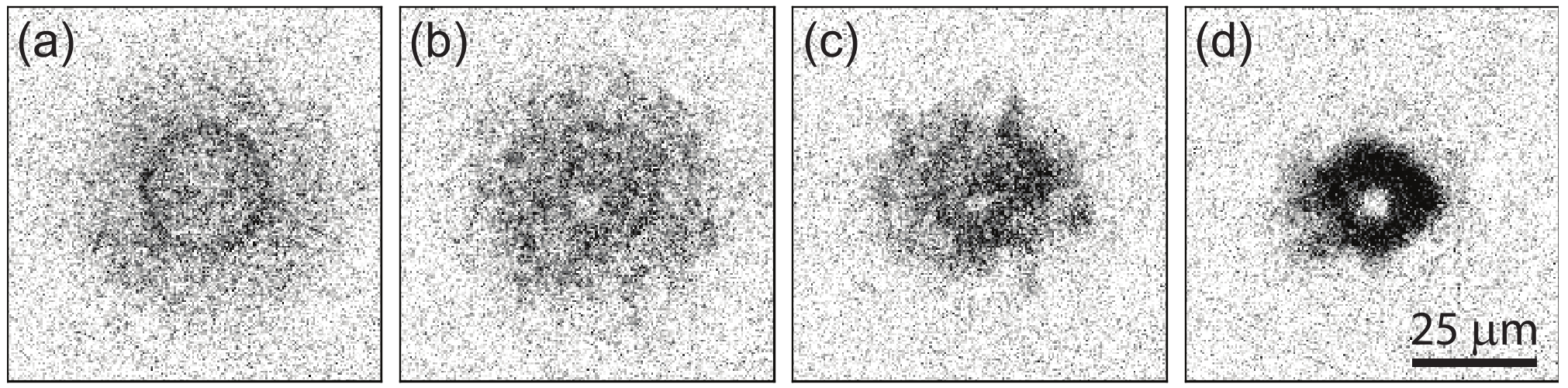}
	\caption{Evolution of a molecular BEC during the last part of the procedure for measuring the current state of the ring. (a) Absorption image showing the vertical column density after relaxing the ring confinement and sweeping the magnetic field from 82.0 to 68.3 mT to lower the interaction energy. (b)-(d) Evolution of the density profile after the optical trap is shut off, for 1.5, 3.5 and 5.5 ms time-of-flight. Radial magnetic lensing improves the signal-to-noise ratio in detecting the vortex core associated with the persistent current. Each image is from a separate realization of the experiment.}
	\label{fig:tof_series}
\end{figure}

It was challenging to adapt the supercurrent detection procedures developed with ring BECs~\cite{RamanathanSuperflowPRL11, Murray2013, EckelInterferometricPRX14, Marti2015, Haug2018, Safaei2019} to rings of strongly interacting light fermionic atoms. Lower condensate fraction in fermionic systems, rapid expansion due to the high chemical potential, and pair breaking in the BCS limit all reduce coherence and the signal to noise ratio in images. These procedures were most effective in the BEC limit after lowering the interaction energy as much as possible. While the field was still at 82.0 mT, we relaxed the radial confinement by lowering the ring beam power to $5\%$ of its initial value over 100 ms, changing the profile of the cloud to that of Fig.~\ref{fig:tof_series}(a). This transformed a current with winding number $\ell$ into $\ell$ singly-charged vortices in the central region that were too small to detect optically. We then swept the magnetic field from 82.0 mT to 68.3 mT in 20 ms, reducing the scattering length by 96$\%$. (Ramping to even lower field caused the three-body loss rate to become too high during the detection procedure.) Next, we turned off the trap and allowed the atoms to evolve for 5.5 ms in a magnetic field with a weak radial curvature. This caused radial focusing, which increased the signal-to-noise ratio in the absorption image taken at the end, using the $\ket{m_J = -1/2,  m_I = 1} \rightarrow \ket{m_J = -3/2,  m_I = 1}$ transition. Fig.~\ref{fig:tof_series} shows typical evolution of the density profile for a system prepared in an $|\ell|=1$ current state. The single hole indicates the presence of a single vortex~\cite{l=2note}.

In this work we initialized the current state by stirring~\cite{BrandGeneratingJoPB01, WrightDrivingPRL13}, but note that phase imprinting is also possible\cite{RyuObservationPRL07,Wright2009, RamanathanSuperflowPRL11, Kumar2018} and was demonstrated with fermions by another group while this paper was in review~\cite{Roatiprivatecomm}. In our system spontaneous currents often appeared during initial formation of the molecular BEC, and stirring allowed deterministic preparation of a selected current state even when the initial current state was uncertain, which is not possible with phase imprinting.  We created a repulsive stirring potential with a steerable blue-detuned beam ($\lambda$ = 635 nm, radius $6(1)$ $\mu$m). To initialize the system in a zero-current state, we kept the beam stationary at one point on the ring, increased the laser power linearly over 100 ms until the peak of the repulsive potential was around $1.5 \mu$, held for 100 ms, then ramped the beam off in 100 ms. After this procedure the probability of detecting a non-zero current was $0.00_{-0.00}^{+0.02}$~ (Uncertainties are $1\sigma$ Bayesian binomial confidence intervals~\cite{CameronEstimationPASA11}).

To create a current we accelerated the stirring beam around the ring at 100 rad/s$^2$ up to a maximum angular velocity that we held constant for 300 ms, then ramped the beam power off linearly in the final 100 ms. The angular frequency of a quantized current of pairs with winding number $\ell$ in our ring was $\ell \Omega_0 \equiv \ell \hbar/(m_{\text{pair}}R^2) = 2\pi \ell \cdot 5.83(2)$ rad$\cdot$s$^{-1}$. The probability of creating an $\ell=1$ current ($P_{\ell = 1}$) became significant for stirring frequencies near 0.5 $\Omega_0$, increasing to $\approx$1 above 0.7 $\Omega_0$. We have created higher current states by stirring at higher angular velocities, but focus here on creation and decay of the $\ell=1$ state.

\begin{figure}[t!]
	\centering
	\includegraphics[width=\columnwidth]{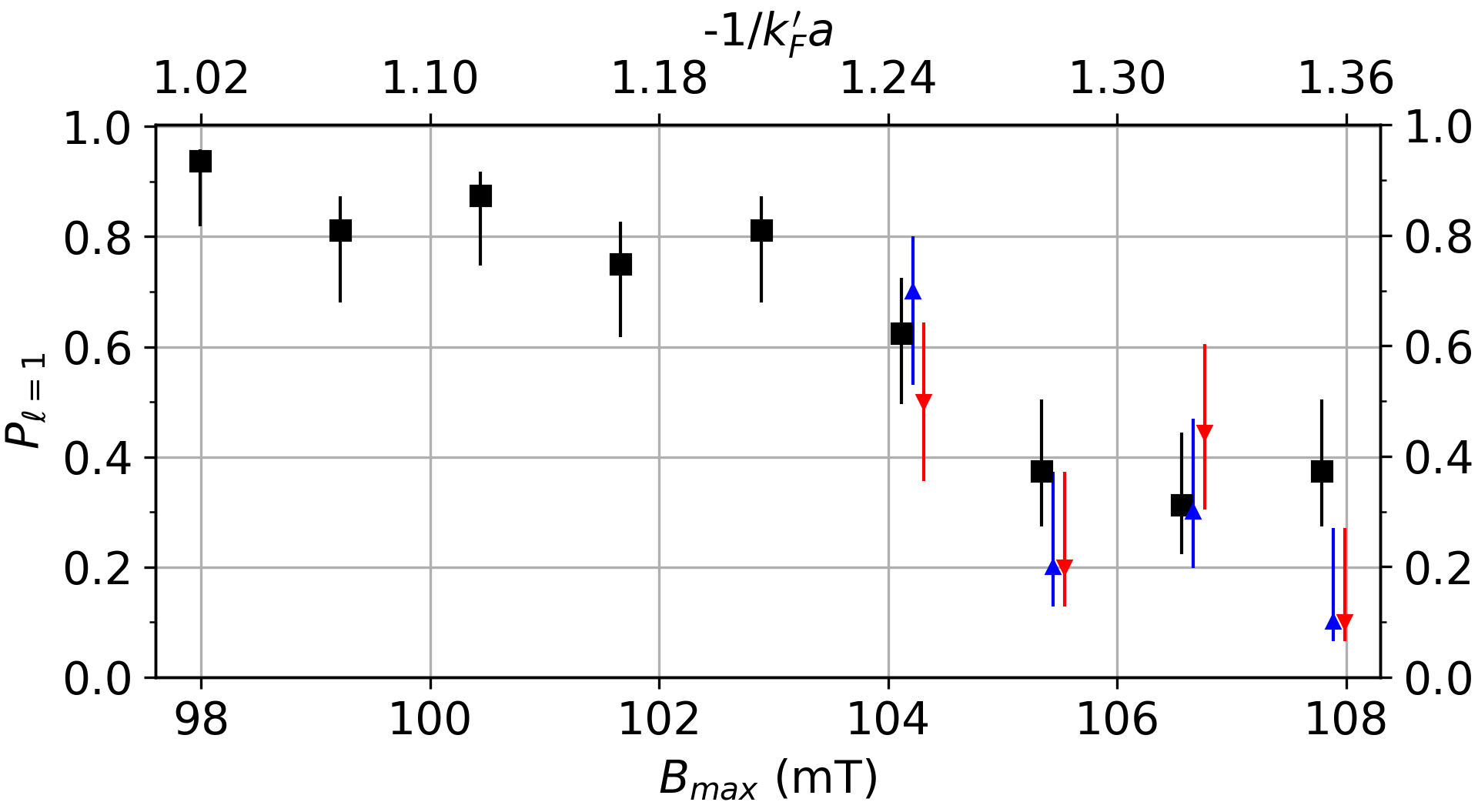}
	\caption{Probability of detecting an $\ell = 1$ current after preparing the superfluid ring in that state near resonance (82 mT) then ramping the interactions into the BCS regime and back. The horizontal axis is the maximum magnetic field ($B_\text{max}$) used in the ramp (lower scale) and the interaction parameter $-1/k_F'a$ (upper scale) where $a$ is the scattering length and $2\pi/k_F'=1.49$ $\mu$m is the local Fermi wavelength at the ``weak point'' of the ring. Black squares represent 16 runs averaged, where $B$ was ramped up and down in 0.2 ms with no hold time at $B_\text{max}$. Triangles are data (10 runs each point) obtained when holding at $B_\text{max}$ for 0.1 s (upright, blue online) and 0.2 s (inverted, red online), and are horizontally offset (from squares) for clarity. Uncertainties are 1$\sigma$ Bayesian binomial confidence intervals~\cite{CameronEstimationPASA11}.}
	\label{fig:survival}
\end{figure}

In our system $T/T_c<0.5$ over a significant range of interaction strengths near resonance, and under these conditions persistent currents survived for up to 10 seconds, limited by losses from background gas collisions ($1/e$ lifetime 12 s) or three-body collisions. (When losses reduced the total atom number below $10^4$ we could no longer distinguish vortices from thermal density fluctuations using the detection procedure described above.) When ramping into the BCS regime $T_c$ falls exponentially, and thermally activated phase slips to lower energy states should occur as $T/T_c \rightarrow1$ at the weakest point of the ring~\cite{Mathey2014, Kumar2017, Kunimi2019}. We estimated the local Fermi temperature at that point to be $T_F'= 0.69(1)$ $\mu$K and the inverse Fermi wavenumber was $1/k_F'=0.24(1)$ $\mu$m. Neglecting small corrections due to trap confinement~\cite{Baranov1998}, the expected critical temperature for the superfluid transition is $T_c \approx 0.277\,T_F e^{-\pi/2k_F\vert a\vert}$~\cite{Gorkov1961}. Given our measurement that $T=25(5)$ nK in the BCS limit, the superfluid density should vanish at the weak point of the ring when $B=106(3)$ mT ($-1/k_F' a = 1.3(1)$).

To characterize the decay of the current around this interaction strength, we prepared the system in the $\ell=1$ current state at $B_i$ = 82.0 mT, swept the magnetic field in 100 ms to a value $B_\text{max}$ in the BCS regime, then swept back to $B_i$ in 100 ms before measuring the final current state. This sweep rate is slow enough to be adiabatic, causing no detectable excitation of collective modes. We found that the current did not decay for $B_\text{max}< 98.0$ mT ($-1/k_F'a < 1$). The data in Figure~\ref{fig:survival} show the decreasing probability of detecting an $\ell=1$ current ($P_{\ell = 1}$) for $B_{\text{max}}$ from 98.0 mT up to 107.8 mT (our technical limit). The decrease of $P_{\ell = 1}$ over the range from 98.0 mT to 105 mT is consistent with expectations that the rate of decay via thermally activated phase slips increases as $T/T_c\rightarrow1$~\cite{McCumberTimePRB70, Mathey2014, Kumar2017, Kunimi2019}. The velocity of the flow (0.44 mm/s) should have a negligible effect on decay of the current since the kinetic energy per pair is less than 0.01 $k_B T$.

Because our detection procedure requires ramping back to the BEC limit, interpretation of the data when any part of the ring is quenched to the normal phase requires consideration of spontaneous current formation, the damping of the normal current, and the effect of thermal phase fluctuations. Spontaneous currents can appear during sufficiently rapid merging of independent superfluid regions~\cite{SchererVortexPRL07, WeilerSpontaneousN08}. This can occur if there is significant azimuthal variation in the ring potential minimum~\cite{Aidelsburger2017}, or via the Kibble-Zurek mechanism during a fast quench to the superfluid phase~\cite{Corman2014, DasWindingSR12}. To determine whether spontaneous current formation was significant for our experimental conditions we prepared the atoms in the $\ell=0$ state and measured the final current state after similar ramps to the BCS regime. The probability of observing $\ell\neq0$ was $0.04_{-0.01}^{+0.08}$, indicating that the non-zero probabilities in Fig.~\ref{fig:survival} can be attributed to initializing the system in the $\ell=1$ current state.

When a normal fluid circulating around a ring is driven into a superfluid phase, it will most likely form in the quantized current state that minimizes the free energy~\cite{Hess1967,Chen2018}. When phase fluctuations are small the distribution of final current states is sharply peaked, with the probability of one state near unity. Large phase fluctuations broaden the distribution and make the result non-deterministic. When $B_{\text{max}}$ is high enough that the ring is broken by a region in the normal phase, damping of the current (and excitations in the remaining superfluid) should cause $P_{\ell = 1}$ to fall to zero eventually. For linear ramps up to $B_{\text{max}}$ and immediately back down, $P_{\ell = 1}$ did not fall to zero and was nearly the same for the highest three values of $B_{\text{max}}$, with an average value of $0.35^{+0.07}_{-0.06}$.

To obtain more information about the time scale for damping we repeated the procedure, adding a hold time of either 0.1 or 0.2 s at the highest values of $B_\text{max}$ (see offset data in Fig.~\ref{fig:survival}). Again, $P_{\ell = 1}$ did not fall to zero, and the dependence on $B_{\text{max}}$ was weak ($0.04\pm0.1$ /mT for the three highest values of $B_{\text{max}}$, similar for 0.1 s and 0.2 s hold).  For a 0.1 s hold, the average value of $P_{\ell = 1}$ for the three highest values of $B_{\text{max}}$ was $P_{\ell = 1}=0.20^{+0.09}_{-0.05}$. For 0.2 s it was $0.24^{+0.09}_{-0.06}$. Fitting to these average values for each hold time, the estimated decay time was 0.5 s, with a 1-$\sigma$ lower bound of 0.25 s. This is longer than our ramp times, and much longer than the few-millisecond timescale for sound to propagate around the ring. We did not systematically investigate longer hold times because heating was non-negligible and we could no longer treat the temperature as nearly constant. The most plausible explanation for the data at the right of Fig.~\ref{fig:survival} is that the average total current remained significantly greater than zero even when part of the ring was driven normal, and thermal phase fluctuations and/or long-wavelength excitations in the superfluid broadened the distribution of final current states after the superfluid ring reconnected. It should be possible to study these current decay and reconnection dynamics in detail in future experiments using interferometric techniques in a ``target" or double-ring trap configuration~\cite{EckelInterferometricPRX14}.

In conclusion, we have studied persistent currents in a fermionic matter-wave ``circuit'' across a range of interaction strengths. We initialized the system in a selected current state and detected single-quantum changes in the current state. We maintained low enough temperatures for supercurrents to survive well into the BCS regime and found that the potential was smooth enough for normal currents to be relatively long-lived. These results also provide a framework enabling future studies of transport and non-equilibrium phenomena in rings of ultracold fermionic atoms. We observed spontaneous currents for faster interaction ramps, indicating an opportunity to study the Kibble-Zurek mechanism with fermions~\cite{Ko2019} in the annular geometry originally proposed by Zurek~\cite{ZurekCosmologicalN85}. In a spin-imbalanced ring of fermionic atoms it may be possible to create $\pi$-Josephson junctions~\cite{Kashimura2011} and search for evidence of unconventional spin-polarized superfluid phases~\cite{Liao2010a, Yanase2009}. Finally, the tight transverse confinement achieved in these experiments could be increased to realize quasi-2D and 1D rings of fermions with tunable interactions, where non-Fermi-liquid behavior is expected and parity effects can be significant~\cite{Pecci2021}.

\begin{acknowledgments}
We thank J. Evans, S. Khatry, L. Bezerra, and A. Woronecki for key technical contributions to the experimental apparatus and recognize D. Adams, C. Grant, and D. Collins for critical engineering support. We also thank R. Onofrio for insightful discussions about ultracold fermionic systems. This work was supported by the NSF (Grant No. PHY-1707557).
\end{acknowledgments}


\begin{thebibliography}{66}
\bibitem [{l=2()}]{l=2note}%
	\BibitemOpen
	\href@noop {} {}\bibinfo {note} {{ For $|\ell|>1$ the multiply charged
			current fragments into singly-charged vortices before the cloud is released,
			with an $|\ell|=2$ current resulting in two distinct holes.}}\BibitemShut
	{Stop}%
\end{thebibliography}
\end{document}


\title{Supplemental Material: Persistent currents in rings of ultracold fermionic atoms: }

\author{Yanping Cai}
\author{Daniel G. Allman}
\author{Parth Sabharwal}
\author{Kevin C. Wright}

\email{kevin.wright@dartmouth.edu}

\affiliation{Department of Physics and Astronomy, Dartmouth College, 6127 Wilder Laboratory, Hanover NH 03766, USA}

\maketitle

\section{Preparing the Degenerate Fermi Gas of $^6$Li}

 We begin our experimental procedure with successive laser cooling of $^6$Li atoms in a 2D magneto-optical trap (MOT) and 3D MOT, followed by $D_1$ grey molasses cooling and optical pumping of the atoms into the $F=1/2$ hyperfine ground states. We capture $10^7$ of these atoms in a 40 Watt 1064 nm crossed-beam optical dipole trap and perform forced evaporative cooling in this trap at 33 mT to obtain a balanced spin mixture of $10^6$ atoms in the $\ket{m_J = -1/2,  m_I = 1}$ and $\ket{m_J = -1/2,  m_I = 0}$ states at a temperature of 30 $\mu$K. We then use a 15 Watt 1064 nm movable optical trap to transport the atoms from the 3DMOT chamber to an octagonal glass cell. We use a matched pair of objectives above and below the cell ($f=30$ mm, NA=0.3) for high-resolution projection and imaging in the vertical direction. Water-cooled magnet coils surrounding the cell generate a nearly uniform magnetic field of more than 0.1 T, which we use for tuning the atomic interactions. The vacuum-limited lifetime of atoms in the cell during these experiments was 12(1) seconds.

After transporting atoms to the cell we transfer them into a ring-dimple trap formed by the intersection of a horizontally propagating ``sheet'' beam and a vertically propagating ring beam. The sheet beam is a $\lambda$=1068 nm (red-detuned) asymmetric Gaussian beam with a horizontal (vertical) waist of 290 $\mu$m (7 $\mu$m). The potential near the focus of this beam is radially symmetric in the horizontal direction, with an asymmetry of less than 3\% at a radius of 100 $\mu$m. At the maximum beam power of 4 Watts the trap depth (for molecules) is 150 $\mu$K, the radial trap frequency is 140 Hz, and the vertical trap frequency is 5.8 kHz. All reported trap frequencies were determined by observations of parametric resonance when modulating the power in the trapping beams.

\begin{figure}[ht]
	\centering
	\includegraphics[width=5in]{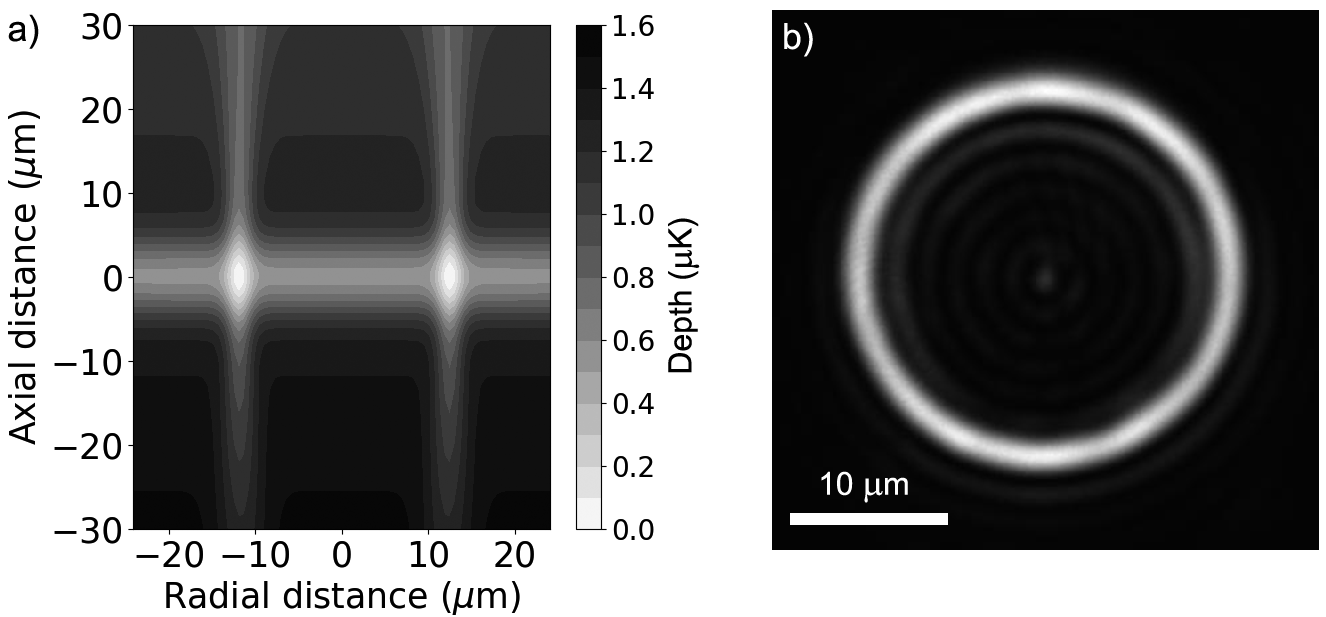}
	\caption{a) Contour plot of an idealized model of the trap potential in the vertical direction through its center. Potential is shown in $\mu$K relative to the lowest point of the ring potential. b) Intensity of the ring-pattern beam as seen when re-imaged onto a camera after it passes through the glass cell. The diffraction fringes are a result of partial apodization of the beam with an adjustable, movable iris that we use to help smooth and control the width of the ring.}
	\label{fig:trap_potential}
\end{figure}

\begin{figure}[h!]
	\centering
	\includegraphics[width=5in]{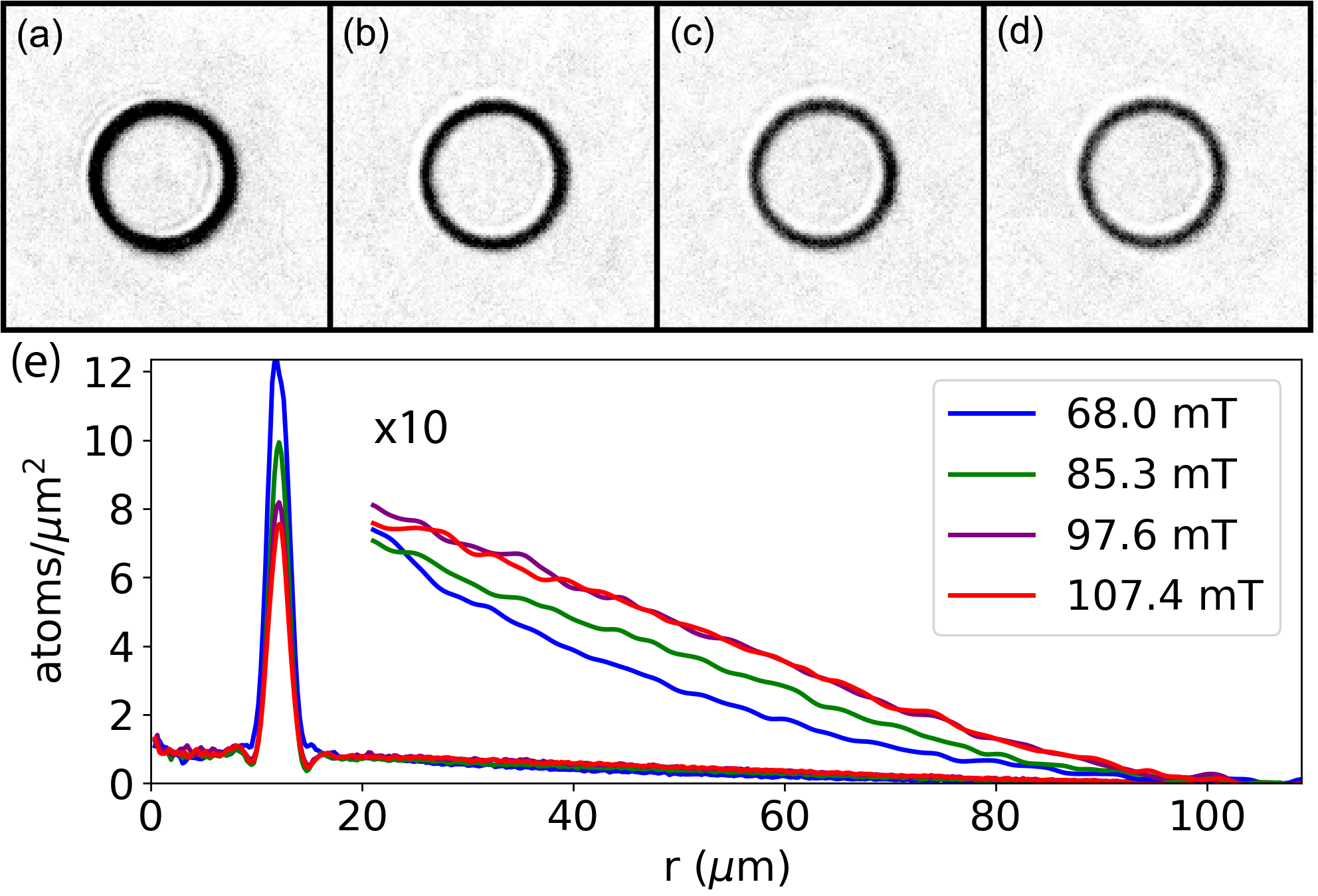}
	\caption{(a)-(d) Absorption images of $1.0(1)\times10^4$ $^6$Li atoms per spin state in the ring-dimple trap showing changes in  the column density distribution when the interactions are tuned from the BEC regime to the BCS regime using a Feshbach resonance at 83.2 mT. The magnetic field for each image is: (a) $B=68.3$ mT (b) $B=85.7$ mT (c) $B=98.0$ mT (d) $B=107.8$ mT. Each image is an average of 10 experimental runs. The field of view is a 63 $\mu$m $\times$ 63 $\mu$m region of a larger image. (e) Azimuthally averaged column density from (a-d) out to larger radii, showing the redistribution of atoms from ring to halo as the interactions are ramped from the BEC to BCS limit. Data for $r>20$ is shown $\times10$ to emphasize the change in the halo population. The data is shown here without correction for blurring due to resolution limits. After correction, the actual peak column densities are higher (See section~\ref{section:density}).}	\label{fig:ring_insitu}
\end{figure}

For these experiments, the vertically propagating ring beam was a $\lambda$=780 nm (red-detuned) laser field with a total power of 0.85 mW. We generated the ring pattern by passing a spatially filtered nearly-Gaussian beam through a positive axicon with a 5$^\circ$ cone angle, followed by an achromatic doublet which brings the beam to a narrow annular focus. After this focus we directed the beam through another 5$^\circ$ positive axicon to make the propagating pattern telecentric before re-imaging onto the atoms using a 1 meter focal length tube lens and the objective lens positioned below the cell. We also used an adjustable iris positioned just after the ring focus behind the second axicon to partially apodize the beam (apodization only at the outer radius). We did this to improve the smoothness of the projected beam and provide finer control over its width and azimuthal symmetry. The incomplete apodization results in diffraction of a small amount of light into the region inside the ring, as shown in Fig.~\ref{fig:trap_potential}(b). The only notable effect is a slight increase in the atomic density at the location of the central Poisson spot (visible in Fig.~\ref{fig:ring_insitu}), and we do not include these fringes in the model of the trap described below.

We determined the average radius of the ring minimum to be 12.0(1) $\mu$m by fitting a cylindrically symmetric 2D ring profile to absorption images (28 averaged) of the atoms in the ring-dimple trap. The dominant contribution to the uncertainty in this measurement is from the magnification of the imaging system. For the data reported in this work the radial trapping frequency of the ring potential was $\omega_r=2 \pi \cdot 4.0(2) \times 10^3$ s$^{-1}$. Given the power in the beam, the measured radial trapping frequency and the assumption that the radial profile is nearly Gaussian, the calculated 1/$e^2$ half-width of the ring is 2.2(1) $\mu$m. The final stage evaporative cooling is performed at 82.0 mT by decreasing the sheet beam power to 40 mW with the ring beam power held constant at 0.85 mW. After evaporation the radial trap frequency for atoms in the sheet is 37(5) Hz and the vertical trap frequency inside (outside) the ring region is is 1.5(1) kHz (1.4(1) kHz).

 In the deep BCS limit the chemical potential, $\mu$, is nearly equal to the Fermi energy, $E_F$, but it decreases as the interactions are ramped across the Feshbach resonance. Fig.~\ref{fig:ring_insitu} shows the changing spatial distribution of atoms in the trap for four different values of the magnetic bias field spanning the BEC and BCS limits. In the weakly-interacting molecular BEC (mBEC) limit at 68.3 mT, shown in Fig.~\ref{fig:ring_insitu}(a), around 30\% of the atoms are in the spatial region of the ring-dimple. The remainder form a broad halo more readily seen in the plot of azimuthally averaged column density (blue line) shown in Fig.~\ref{fig:ring_insitu}(e). In the (weakly-attractive) BCS limit, shown in Fig.~\ref{fig:ring_insitu}(d), the increase of the chemical potential to the Fermi energy ($E_F$) reduces the fraction of atoms in the ring-dimple spatial region to less than 20\%, with the balance moving to the halo as shown by the red line in Fig.~\ref{fig:ring_insitu}(e).

\section{Fermi Energy and Chemical Potential}

\begin{figure}[!ht]
	\centering
	\includegraphics[width=4in]{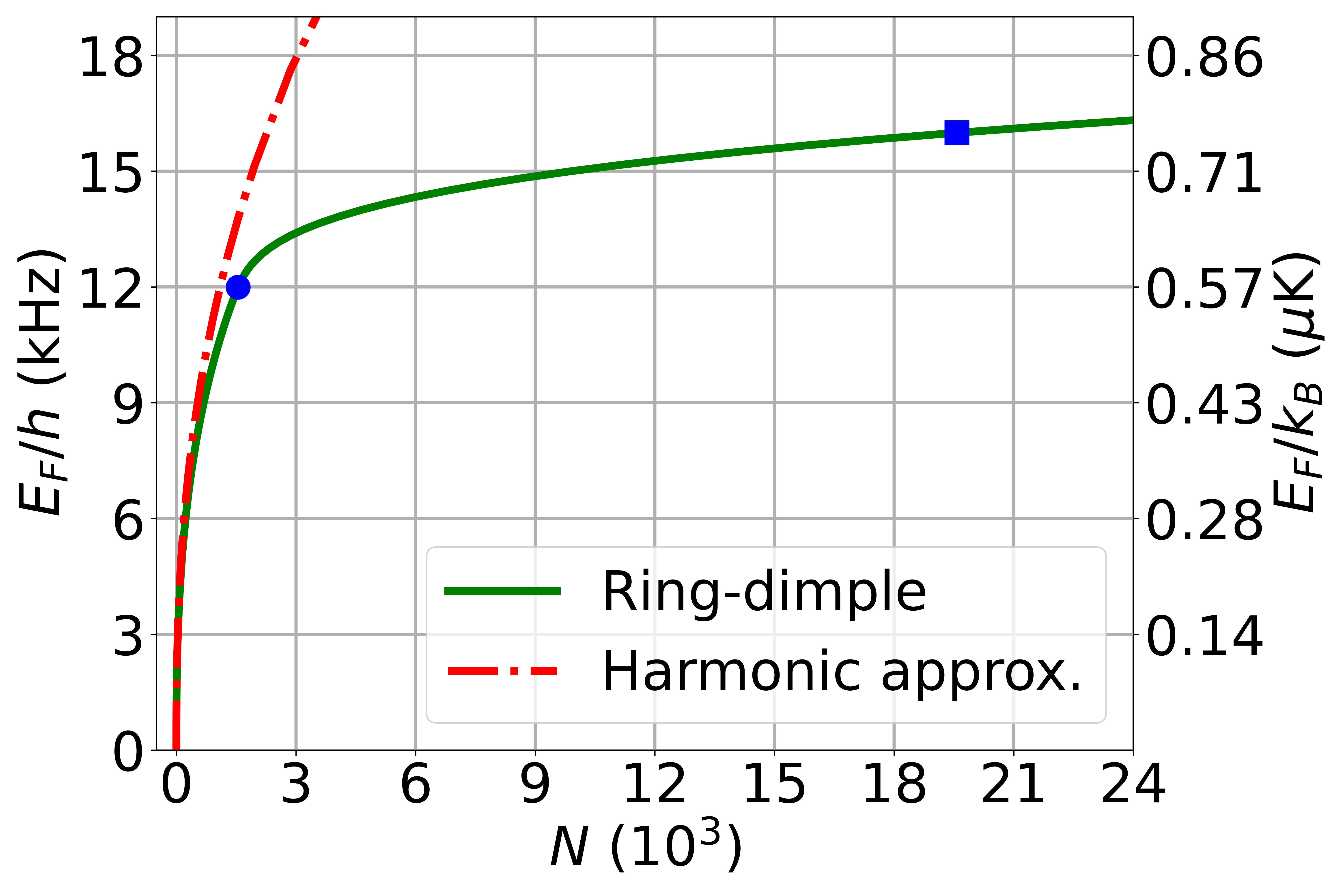}
	\caption{Calculated Fermi energy $E_F$ versus total atom number $N_{\text{tot}}$ for the trap parameters used in our experiment obtained using the model density of states, equation \eqref{DoS3D}. The circle marks the point where $N_{\text{tot}}$ exceeds the capacity of the ring dimple, and the square is the point corresponding to the  Fermi energy for the number of trapped atoms in our experiment ($N_{\text{tot}},\ E_F$). The dash-dotted line shows $E_F$ for a harmonic approximation to the ring potential minimum, for comparison.}
	\label{fig:FermiEnergy}
\end{figure}

To calculate the Fermi energy ($E_F$) of the system in the BCS limit we computed the total (spin up \textit{and} down) semiclassical density of states in 3D,
\begin{equation}\label{DoS3D}
    g_{3D}(E)=2\times\frac{4\pi m}{(2\pi\hbar)^3}\int_{V(\textbf{r})\leq E}d^3r\sqrt{2m(E-V(\textbf{r}))},
\end{equation}
 where $V(\textbf{r})$ is a model of the trap incorporating the potential of the sheet beam, the ring beam, and gravity. (Quasi-2D confinement at the edge of the halo has a negligible effect on $E_F$, but not on temperature measurements described below.) From the density of states we then used the defining relation $N_{\text{tot}}=\int_0^{E_F}g_{3D}(E)dE$ to numerically compute the Fermi energy $E_F(N_{\text{tot}})$. We modelled the ring beam as having a ring-shaped Gaussian intensity profile in its plane of best focus, and numerically propagated it through focus using the angular spectrum method to obtain its full 3D profile. The sheet beam was taken to be an asymmetric Gaussian beam, whose propagation has a known analytic form. Furthermore, to make computations more tractable, we assumed azimuthal symmetry of the potential, since the sheet beam ellipticity is $<3\ \%$ in the region of interest. For $1\times10^4$ atoms in each spin state, the estimated Fermi temperature ($T_F$) is 0.77(6) $\mu$K. Uncertainty in $E_F$ is due in most part to our uncertainty in the ring width. This is because the dimple capacity depends quite sensitively on the ring width, which in turn affects the estimate of  $E_F$.

Fig.~\ref{fig:FermiEnergy} shows a plot of $E_F(N)$ for the trap configuration used in our experiment. The ``ring-dimple'' nature of the trap is evident in the rapid increase of $E_F(N)$ up to the dimple capacity of about 1500 atoms, followed by a much slower increase in $E_F$ as atoms begin to populate states that spread out into the sheet potential. Here we emphasize that the population of atoms in the dimple, defined via energy considerations, is smaller than the apparent population in the ring-shaped region of higher density seen in absorption images. The red dash-dotted line in Fig.~\ref{fig:FermiEnergy} is $E_F(N)$ for a ring of the same radius with idealized harmonic confinement in the transverse direction, given by the expression
\begin{equation}
    \frac{E_F}{\hbar\Omega_0}=\left(\frac{15N_{\text{tot}}}{16}\right)^{2/5}\left(\frac{\Bar{\omega}}{\Omega_0}\right)^{4/5}
\end{equation}
where $\Bar{\omega} = \sqrt{\omega_z \omega_r}$ is the the geometric mean of the trapping frequencies, and the quantized rotation frequency is $\Omega_0\equiv\hbar/(2mR^2)$. This approximation is reasonably valid for many previous experiments involving ring BECs, but is not appropriate for our experimental conditions.

\section{Temperature Measurements}

The temperature can be determined from the density profile of the halo and knowledge of the trapping potential for $r>20$ $\mu$m. In this region, we can approximate the potential by $V(r,z)=V_{0,s} + \frac{m}{2}(\omega_s^2r^2+\omega_z^2z^2)$ where $V_{0,s}$ is the sheet-only potential at the origin, $\omega_s$ is the radial trap frequency of the sheet, and $\omega_z$ is the vertical trap frequency in the sheet. It is important to note that for deeply degenerate Fermi gases, absolute temperature enters into the fit of the data only in the far, dilute wings of the density distribution where the system is potentially quasi-2D. We allow for mixed dimensionality in our description of the density by treating the vertical energies quantum mechanically and the radial energies semi-classically. This approach is similar to those used to model quantum wells in solid state systems, except that our tight vertical confinement is harmonic, not hard-walled. In this way, we may write a hybrid description of the density of states
\begin{equation}\label{hybridDoS}
    g_j(E)=\frac{s}{(2\pi\hbar)^2}\iint d^2rd^2p\ \delta\left(E-\frac{|\textbf{p}|^2}{2m}-\hbar\omega_zj-V_r(r)\right)
\end{equation}
which represents the density of available states in the $j^{th}$ axial harmonic oscillator level ($j=0,1,...)$, for a system with $s$ spin degrees of freedom. We have defined $V_r(r)=V_{0,s}+\hbar\omega_z/2+m\omega_s^2r^2/2$, explicitly accounting for the finite zero-point energy of the axial motion. Integrating over momenta and summing over $j$, we identify the local density of states 
\begin{equation}\label{localHybridDoS}
    g(r;E)=s\frac{m}{2\pi\hbar^2}\lceil\epsilon(r)\rceil\Theta(\epsilon(r))
\end{equation}
where $\epsilon(r)\equiv\frac{E-V_r(r)}{\hbar\omega_z}$, $\lceil x\rceil$ is the ceiling function, and $\Theta$ is the Heaviside step function. The column density $n_{2D}(r)$ is found by integrating the Fermi-Dirac-weighted local density of states over $E$, giving
\begin{equation}\label{ncol}
n_{2D}(r)=\frac{s}{\lambda_T^2}\int_0^{\infty}dx\frac{\lceil x/\eta\rceil}{e^{x-\tilde{\mu}(r)}+1} 
=\frac{s}{\lambda_T^2}\sum_{j=0}^{\infty}F_0(\tilde{\mu}(r)-j\eta)
\end{equation}
where we have further defined $\lambda_T^2\equiv\frac{2\pi\hbar^2}{mk_BT}$,  $\eta\equiv\frac{\hbar\omega_z}{k_BT}$, $\tilde{\mu}(r)\equiv\frac{\mu-V(r)}{k_BT}$, and $F_0(x)=\log(1+e^x)$, which is a special case of a Fermi-Dirac integral, $F_{\nu}(x)$, of order $\nu$. The chemical potential $\mu$ is measured above $V_{0,s}+\hbar\omega_z/2$. The integral expression looks remarkably similar to the order $1$ Fermi-Dirac integral, except for the presence of the ceiling function in the integrand, which accounts for the discrete axial energy levels. This discreteness is blurred out if either $\eta$ or $\eta/\tilde{\mu}(r)$ is small, which corresponds to the 3D limit. In this case, we can replace $\lceil x/\eta\rceil$ with $x/\eta$, and the resulting expression gives the proper integrated 3D column density
\begin{equation}\label{3Dlimit}
    n_{2D}(r)\approx \frac{s}{\eta\lambda_T^2}F_1(\tilde{\mu}(r))\ ; \ \ \ \ \ \eta\ll 1\ \text{or}\ \eta/\tilde{\mu}(r)\ll 1
\end{equation}
Conversely, if $\eta\gg 1$ and $\eta/\tilde{\mu}(r)\gg 1$, we approach the $2D$ limit, and we may replace $\lceil x/\eta\rceil$ with $1$, and the resulting column density gives the proper $2D$ density
\begin{equation}\label{2Dlimit}
    n_{2D}(r)\approx\frac{s}{\lambda_T^2}F_0(\tilde{\mu}(r))\ ; \ \ \ \ \ \eta\gg 1\ \text{and}\ \eta/\tilde{\mu}(r)\gg 1
\end{equation}
We may write $\tilde{\mu}(r;\tilde{\mu}_0,r_F)=\tilde{\mu}_0(1-r^2/r_F^2)$, with $r_F^2\equiv\frac{2\mu}{m\omega_s^2}$ and $\tilde{\mu}_0\equiv \frac{\mu}{k_BT}$, and define our column density fit function with fit parameters $\boldsymbol{\alpha}=(r_F,n_\text{max},\tilde{\mu}_0)$ 
\begin{equation}\label{fullfit}
    n_\text{fit}(r,\boldsymbol{\alpha})= \frac{n_\text{max}}{C}\int_0^{\infty}dx\frac{\lceil x/\eta\rceil}{e^{x-\tilde{\mu}(r;\tilde{\mu}_0,r_F)}+1}  
    =\frac{n_\text{max}}{C}\sum_{j=0}^{\infty}F_0(\tilde{\mu}(r;\tilde{\mu}_0,r_F)-j\eta) 
\end{equation}
where $C$ is the peak value of the integral or sum, i.e. at $r=0$, and depends on the other fit parameters. $n_\text{max}$ therefore represents the peak density. One may truncate the sum whenever the argument becomes large and negative, which for $\tilde{\mu},\eta\lesssim10$ requires perhaps $20$ terms, ($F_{\nu}(x)\sim e^{x}$ for $x\ll0$) making it useful for fitting. 

Finally, we may eliminate $\eta$ as a fit parameter as $\eta=\frac{\hbar\omega_z}{k_BT}=\frac{\hbar\omega_z\tilde{\mu}_0}{\mu}=\frac{2\hbar\omega_z\tilde{\mu}_0}{m\omega_s^2r_F^2}$ as long as $\tilde{\mu}_0>0$. We pay a modest price for this parameter elimination by introducing uncertainty in $\boldsymbol{\alpha}$ via the vertical and radial sheet trap frequencies $\omega_z$ and $\omega_s$. The largest source of temperature uncertainty is due to uncertainty in $\omega_s$, which by itself introduces several nK uncertainty on our reported temperature in the BCS limit. 

The temperature of the gas in the molecular BEC limit can be determined in by a similar fit to the halo radial density profile. At 68.3 mT the profile for $r>20$ $\mu$m is well approximated by a Gaussian, which is expected for a thermal gas in a harmonic potential. The radius of the fit is 68(1) $\mu$m, indicating a temperature of 90(3) nK. Since the gas is non-degenerate and the estimated temperature is just above the level spacing in the vertical direction $\hbar \omega_z / k_B = 70$ nK we checked this estimate against one obtained by measuring the rate of ballistic expansion on the halo in the vertical direction. Fitting a series of images taken from a horizontal view, we obtained an estimate of 70(20) nK, which is in reasonable agreement with that obtained from the Gaussian fit to the radial density profile.

\section{2D and 3D Number Density}\label{section:density}
We obtained and cross-checked two estimates of the peak 3D density of atoms in the ring in the BCS limit. The first estimate was to make the local density approximation and relate the global Fermi energy $E_F$ calculated from the total atom number and trap potential (see above) to the peak 3D density $n_{3D}$ via the $T=0$ ideal (scattering length $a=0$) fermion equation of state $E_F = (\hbar^2/2 m)(3\pi^2n_{3D})^{2/3}$, which is accurate to $\mathcal{O}((T/T_F)^2)$. The Fermi temperature of 0.77 $\mu$K obtained as described above corresponds to a local density of 2.8 atoms/$\mu$m$^3$ in the BCS limit. This model also predicts the 3D  distribution of atoms in the trap,  $n_{3D}(\textbf{r})=\frac{1}{3\pi^2}\left(\frac{2m}{\hbar^2}\right)^{3/2}\left(E_F-V(\textbf{r})\right)^{3/2}$. Integrating this along the (vertical) imaging direction gives an estimate of the 2D column density as it would appear with perfect imaging resolution. Convolving this calculated 2D density profile with a model of our imaging point spread function (a 1.7 $\mu$m airy disk) gives a 2D density profile with a peak value and ring width that are in good agreement with those obtained from in-situ absorption images of atoms taken in the BCS regime (at 107.8 mT). 

The variation in density around the ring minimum is large enough to impact estimates of the lowest value of the magnetic field for which $T/T_c=1$ at some point around the ring. To quantify this we sampled the column density in the absorption images at 200 angular positions around the minimum of the ring potential and applied a multiplicative correction factor to account for the effects of finite-resolution blurring described above. After that correction we found that the 3D density at the ``weak point'' of the ring was 85$\%$ of the average 3D density at the ring minimum, and the local Fermi energy was 90$\%$ of the average value around the ring. We used this estimate to determine the critical interaction strength and magnetic field for which $T/T_c=1$ at that location around the ring. The uncertainty in those calculated values  is dominated by uncertainty in the temperature, however, not by the uncertainty in this estimate of the local Fermi energy.